%% file: main.tex
\documentclass[a4paper]{article}
\usepackage[english]{babel}
\usepackage[utf8]{inputenc}
\usepackage{csquotes}

\usepackage[a4paper,top=2cm,bottom=2cm,left=3cm,right=3cm,marginparwidth=3cm]{geometry}

\usepackage{amsmath}
\usepackage{graphicx}
\usepackage{xcolor}
\usepackage{xargs}
\usepackage[colorlinks=true, allcolors=blue]{hyperref}
\usepackage{multicol}
\usepackage{enumitem}
\usepackage{caption}
\usepackage{authblk}
\usepackage{booktabs}

\usepackage[backend=biber,style=numeric,autocite=plain,sorting=none]{biblatex}
\usepackage{lipsum}
\usepackage{orcidlink}

\begin{document}
\begin{titlepage}


\title{ \bf Decentralized Peer Review in Open Science:\\ A Mechanism Proposal}

\author[1]{Andreas Finke \orcidlink{0000-0002-8237-5046}}
\affil[1]{
D\'epartement de Physique Th\'eorique and Center for Astroparticle Physics, Universit\'e de Gen\`eve, 24 quai Ansermet, 1211 Gen\`eve 4, CH}

\author[2]{Thomas A. Hensel \orcidlink{0000-0002-7497-3329}}
\affil[2]{Department of NanoBiophotonics, Max Planck Institute for Multidisciplinary Sciences, Am Fassberg 11, G{\"o}ttingen, DE}

{
    \makeatletter
    \renewcommand\AB@affilsepx{: \protect\Affilfont}
    \makeatother

    \affil[ ]{Email}

    \makeatletter
    \renewcommand\AB@affilsepx{, \protect\Affilfont}
    \makeatother

    \affil[1]{af@staedy.co}
    \affil[2]{thomas.hensel@mpinat.mpg.de}
}

\maketitle

\begin{abstract}
Peer review is a laborious, yet essential, part of academic publishing with crucial impact on the scientific endeavor. 
The current lack of incentives and transparency harms the credibility of this process.
Researchers are neither rewarded for superior nor penalized for bad reviews.
Additionally, confidential reports cause a loss of insights and make the review process vulnerable to scientific misconduct.
We propose a community-owned and -governed system that 1) remunerates reviewers for their efforts, 2) publishes the (anonymized) reports for scrutiny by the community, 3) tracks reputation of reviewers and 4) provides digital certificates.
Automated by transparent smart-contract blockchain technology, the system aims to increase quality and speed of peer review while lowering the chance and impact of erroneous judgements.
\end{abstract}

\tableofcontents
\end{titlepage}

\newpage

\input{SECTIONS/motivation}

\input{SECTIONS/decentralized-review}

\input{SECTIONS/scientific-impact}

\input{SECTIONS/tokenomics}

\input{SECTIONS/conclusion}
\input{SECTIONS/ack}

\newpage
\input{SECTIONS/appendix}
\printbibliography

\end{document}

%% file: SECTIONS/motivation.tex
\section{Introduction}
Researchers depend on the present system of academic publishing in order to gain visibility and secure funding and positions in a competitive environment.
Number of papers, citations and impact factors are metrics used by administrators and scientists outside of their respective field of expertise to distribute a limited number of grants and jobs among many applicants. Assuming that optimal scientific progress requires allocating resources to the best scientists, publication metrics ought to be based on merit and scientific integrity. In particular, the best young scientists need to be identified and supported before they decide to leave academia in a process that only depends on their work.

Only highest-effort peer review can judge the quality of the output of a scientist timely enough for the purpose of resource allocation. While experts could judge the quality of work and potential of scientists early in their career independently of peer review, they may be too busy to come across or even carefully study work by unknown authors, especially in popular fields. The process of propagating their opinions is also prone to error and manipulation. Conversely, another issue addressed by functional peer review is that of unbiased evaluation of the current output of established scientists in the light of new progress, as well as any changes of interest or work ethic, or simply increasing age.  

The present process of academic publishing is under scrutiny of all stakeholders - and criticized by many.
Rent-seeking of established science journals leaves no room to incentivize high-quality peer review. From the perspective of most journals, the contributors involved (authors, editors and reviewers) work for free\footnote{Do the contributors themselves think that they work for free? Some, often tenured, academics feel either entirely idealistic or understand part of their salary as the reward for their peer review work. Here we advocate for a more egalitarian system that does not require idealism to function and is truly open to all people, independent of their background and work contracts.}, while the journals receive money that society intended to spend on research. Consequently, the quality of reviews decreases, which compromises falsifiability and reproducability of results~\cite{Ioannidis2005}.

Those problems are widely known, and previous attempts at improving the situation have mostly failed.
Free pre-print web services such as \emph{arXiv} enable sharing of manuscripts and claiming priority of an idea.
However, without peer review, important articles may drown in the noise of submissions, while others have undeserved impact.
This results in a state where group dynamics instead of truth governs the evolution of science. Furthermore, pre-prints are incompatible with fully blinded peer review, since author names can be found by reviewers searching a submission's pre-print.

The open access movement promised free retrieval of peer-reviewed publications - at massive costs for institutions and authors alike\footnote{``Double dipping'' is a common practice of hybrid journals to offer institutions subscriptions for access while also offering authors to pay publication fees such that their own article becomes open-access.}~\cite{Brainard2021}. Yet, it did not lead to faster, more transparent review practices or incentives for high-quality review \cite{Lerdau2022,Zhong2022}.
Lately, efforts such as \emph{SciPost}, \emph{eLife} or \emph{PLOS ONE} \cite{SciPost,PLOS,Brainard2022} have started to serve the interests of the respective scientific community as (centralized) online journals.
While these endeavours set high standards for transparency and open access, they do not solve the problem of missing incentives in a system that is no longer driven by idealism.

New technologies are emerging -- commonly referred to as ``web3'' -- that are suited to build community-owned entities with low organizational and legal overhead, so-called Decentralized Autonomous Coorporations/Organizations, or DACs/DAOs.
Blockchains can execute code (``smart contracts'') in a trustless, distributed and transparent fashion as well as keep a record of user assets, be they monetary or scientific\footnote{Such as timestamped proofs of submission or reputation tokens.}, without needing an intermediary.

Here, we describe how aligned monetary and reputation incentives can enable peer review in a credible-neutral way~\cite{credible-neutral} to identify good science with long-term impact beyond immediate citations of an article.

We share this view with a rapidly growing community of blockchain-affine scientists and software builders~\cite{Hamburg2021,Trov2021}. Decentralized science (DeSci) is a global movement in analogy to decentralized finance (DeFi), and the increasing amount of popular conferences organized on the topic suggests that it may have crossed a point of no return\footnote{See \autoref{sec:appendix} for an overview of previous/current attempts for addressing peer review.}.

This paper is structured as follows. We describe the core mechanisms of the decentralized peer review system in \autoref{sec:review}. A crucial component of such a system is the final judgement by reviewers leading to the binary acceptance/rejection decision; in \autoref{sec:paperquality}, we provide more detail on how one may reduce subjectivity in this step.
Economic aspects of the project are discussed in \autoref{sec:tokenomics} before we conclude in \autoref{sec:conclusion}.

%% file: SECTIONS/decentralized-review.tex
\section{Decentralized peer review}
\label{sec:review}
In this section we describe possible mechanisms for all required building blocks of a decentralized peer-review system. 
These are:
\begin{multicols}{2}
\begin{enumerate}[label=\upshape(\arabic*),ref=(\arabic*)]
\label{list:steps}
\item\label{th:assess_paper} Pre-selecting papers 
\item\label{th:allocate_reviewer} Allocating reviewers
\item\label{th:conduct_review} Conducting the review 
\item\label{th:aggregate_eval} Acceptance decision, conflict resolution
\item\label{th:distribute} Distribution of author assets
\item\label{th:evaluate} Evaluation of reviewers
\end{enumerate}
\end{multicols}

Note that an editor traditionally has responsibilities \ref{th:assess_paper}, \ref{th:allocate_reviewer} and \ref{th:aggregate_eval} of the above list.
In our proposed system, tasks \ref{th:assess_paper} and \ref{th:allocate_reviewer} are handled by the community and 
\ref{th:aggregate_eval} is a smart contract deciding based on scores $Q$ submitted by each reviewer. In turn, the collection of these ratings takes place at completion of the review process \ref{th:conduct_review}, which is based on a public (but anonymized) discussion phase with the authors.
In the following section~\ref{sec:paperquality}, we present an information-theoretical measure of paper quality, turning the reviewers' impression of the authors' key achievements into a single number $Q$.

To achieve careful and honest reviews, a combination of payments and reputation impact comes into play.
Reviewers are expected to (and hopefully motivated to) invest effort and time due to their remuneration; they are also expected to act honestly. They are being held accountable post review in step~\ref{th:evaluate} via a reputation system.
Reviews deemed high-quality by users boost the reputation of a reviewer, while unconvincing reviews can significantly lower it. 

If the system is successful, high reputation means high status, which should drive behavior; early participants would likely expect a nonzero chance of success of the system and already be status-seeking. 
Furthermore, step \ref{th:allocate_reviewer} is such that high reputation increases the chance of reviewing allocation. Finally, reputation allows to receive a share of the protocol income. 

Reputation can also be gained (or lost) by participation in step~\ref{th:assess_paper}, and we construct the updates below such that they cannot be easily gamed.
Note that within current centralized peer review, points~\ref{th:distribute} and \ref{th:evaluate} do not have a counterpart. 

The improvements to  \ref{th:assess_paper}, \ref{th:allocate_reviewer}, the transparency of part \ref{th:aggregate_eval}, the open discussion of the paper in~\ref{th:conduct_review}, the access to reviews by scientists building on the work of the paper as required by \ref{th:evaluate}, as well as author benefits such as a proof of submission without revealing their identity or tradeable assets proving acceptance distributed by the system~\ref{th:distribute} are all immediate advantages that can help find product-market-fit and establish the system quickly\footnote{No rights are taken from the authors, who may still submit their work to traditional journals.}.

In the following, we flesh out more details of these steps.
For the sake of effective communication, we occasionally assign values to parameters, stressing that they can all be updated by community governance, enabling flexible fine-tuning for the running system (see section \ref{subsec:governance} for how we envision governance).

\subsection{Pre-review}
This section is concerned with steps \ref{th:assess_paper} and \ref{th:allocate_reviewer} of the above process. 

Naturally, the same expertise that one needs for insightful reviews is also useful for the initial screening of submissions a step that may lead to early rejection of a submission\footnote{Note that it is crucial to filter out low-quality work before the costly review process to mitigate a spam attack vector.}.
This motivates merging early rejections and reviewer allocation into one poll for each new submission.

In detail, we propose the following mechanism for the pre-review process.

\begin{enumerate}
    \item A paper is submitted.
    \item Within at least a time window $T_\text{vote}$ or at most until $T_\text{max}$,
    a voting period allows reviewers to vote once among the options:
    \begin{enumerate}
        \item no opinion \label{enum:noopinion}
        \item flawed or uninteresting, reject immediately \label{enum:reject}
        \item seems interesting \label{enum:interesting}
        \item seems interesting and would review. \label{enum:wantreview}
    \end{enumerate}
    We propose $T_\text{vote} = 72$h and $T_\text{max} = 30$d. 
    \item Let time $t$ be the time after submission. As soon as $t>T_\text{vote}$ \emph{and} over $N_\text{votes}$ are collected, if the number of votes for~(\ref{enum:reject}) exceeds those for~(\ref{enum:interesting}) and~(\ref{enum:wantreview}) combined, the paper is immediately rejected and a minimum submission fee is not refunded.  This stops the process.
    \item If $t \ge T_\text{max}$ and either 
    less than $N_\text{votes}$ votes are collected in total or less than $N_\text{reviews}=3$ votes amass for~(\ref{enum:wantreview}) the paper review process fails and is stopped. In this case, all prepaid costs are automatically refunded to the submitter. 
    \item Else, we have reached the necessary number of votes and reviewer volunteers at some $t_0$ with $T_\text{vote} \le t_0 < T_\text{max}$. The voting period closes at $t=t_0$, but never before $t=T_\text{vote}$. The fractional remaining time of a voter of option~(\ref{enum:wantreview}) at time $t$ given by $\min(1, 1-(t-T_\text{vote})/(T_\text{max}-T_\text{vote}))$, which is linearly decreasing from $1$ to $0$ for $t\in [T_\text{vote}, T_\text{max}]$, is recorded and will suppress their final review remuneration to incentivize voting within $t<T_\text{vote}$.
    \item If more than $N_\text{reviews}$ volunteered for review by voting~(\ref{enum:wantreview}),  $N_\text{reviews}$ accounts are selected randomly from a distribution informed by reputation.\footnote{A natural proposal is to choose probabilities proportional to the reputation taken to a power such as $\alpha = 1/2$, where $\alpha$ is another parameter that can be tuned by governance vote. }
    \item The paper is reviewed and reputation scores are adjusted.
\end{enumerate}

Reputation can be earned  (and lost) already at this stage. For example, we can give a moderate reward of $1$ reputation point to those that voted for~(\ref{enum:reject}) in the case the paper does get immediately rejected. It should, however, not be possible to gain reputation by \emph{always} voting with any option. Lack of precision in the case that a paper they claim has obvious flaws ends up \emph{passing peer review with acceptance} must come with a sufficient cost to disable the strategy of always voting~(\ref{enum:reject}). Let $f_\text{acc/rej}$ the (measured) fractions of fully accepted and early rejected papers, respectively. The reputation loss in the mentioned case should be larger than $f_\text{rej}/f_\text{acc}$. This is the threshold that cancels the gain from correctly predicting those that are indeed immediately rejected ($f_\text{rej} \times 1 - f_\text{acc} \times f_\text{rej}/f_\text{acc} = 0$), assuming we do not update reputation for these voters in the third case where a paper enters review but does get rejected eventually. The prospective large loss of reputation for critical early voters in case of later paper acceptance also incentivizes experts who spotted crucial flaws to communicate these in the open forum (see below, \autoref{sec:reviewing-mechanics}).

For voting with no opinion, option~(\ref{enum:noopinion}), similar considerations apply. The purpose of this option is to reward users engaged in the screening process even if they are not particularly knowledgeable. Since it can be relatively easy to spot the weakest submissions, in case of the outcome of immediate rejection, such indifferent voters may be punished with loss of $ (1-f_\text{rej})/f_\text{rej}$ reputation points. However, when the paper enters review, these indifferent voters can be rewarded with $1$ reputation point. This enables non-experts to gain reputation by identifying obvious flaws of a fraction $f_\text{bad}$ of the fraction $f_\text{rej}$ of immediately rejected papers and otherwise voting with~(\ref{enum:noopinion}): 
The expected win for this strategy, considering cases of correct prediction of immediate rejections, missed identification of immediate rejections, and conservatively voting~(\ref{enum:noopinion}) on all papers that do enter review, respectively, is then
\begin{align} 
  f_\text{rej}\left(f_\text{bad} - (1-f_\text{bad}) \frac{1-f_\text{rej}}{f_\text{rej}} \right)  + (1-f_\text{rej}) = f_{bad} < 1
 \end{align}
and grows simply with the sensitivity (recall) of that voter. 

Finally, the reputation updates for voters of the latter two options can be chosen as follows. If they were wrong, that is, the paper is rejected early, a loss of $1/f_\text{rej} > 1$ compensates a win of $1/(1-f_\text{rej}) > 1$ if the paper does enter review. However, since overlooking substantial flaws is fatal for reviewing experts, those that volunteer for reviewing~(\ref{enum:wantreview}) should, in case of early rejection, have their loss amplified by a factor $>1$ to offset rent-seeking behavior of potential reviewers.

\autoref{tab:reputation-gain} displays the different rewards for the voting under consideration of the eventual turn-out.
The expected reputation change for always voting with any answer are all zero or negative.\footnote{This follows from noting the probabilities for the columns are $f_\text{rej}$, $f_\text{acc}$, $1-f_\text{rej}- f_\text{acc}$, respectively.}  Linearity of the expectation value implies that no fixed or randomized mixture of these strategies a) to d) brings about a positive gain. Hence, voting strategies uncorrelated to paper content fail, as opposed to strategies that indeed add information to the pre-review stage from experienced/knowledgeable users.

\begin{table}[h!]
\centering
\caption{Reputation gain/loss with respect to voting option and eventual turn-out.}
\label{tab:reputation-gain}
\begin{tabular}{l|c|l|l} 
\toprule
option\textbackslash{}outcome & \multicolumn{1}{l|}{reject immediately}    & review and accept                                 & review and reject                                                \\
a: no opinion                             & $-(1-f_\text{rej})/f_\text{rej}$  & $1$  & $1$ \\                                                                                  
b: reject immediately                             & $1$                          & $< -f_\text{rej}/f_\text{acc}$ & $0$  \\
c: interesting                          & $-1/f_\text{rej}$                        & $1/(1-f_\text{rej}) $  & $1/(1-f_\text{rej}) $     \\
d: interesting and want to review & $ < -1/f_\text{rej}$                        & $1/(1-f_\text{rej}) $  & $1/(1-f_\text{rej}) $     \\ 
\bottomrule
\end{tabular}
\end{table}

\subsection{The review process}\label{sec:reviewing-mechanics}
This section is concerned with steps  \ref{th:conduct_review} and \ref{th:aggregate_eval} of the process described at the beginning of \autoref{list:steps}.
The review process is designed to allow for communication between authors, reviewers and the public while preserving anonymity in crucial places.

With the commencement of the review process, a public discussion forum opens specifically for the new submission. The review process is double-blind (author names and reviewer names, as well as their blockchain addresses, are hidden). However, the identities of all \emph{additional} participants in the discussion are public.
The purpose of this forum is not for authors and reviewers to interact freely, but rather to allow the community to follow along and to point out flaws or merits of the paper. This is of particular importance in case these have not already been noticed by the reviewers. Some users may have based voting decisions at the previous prediction stage on particularly insightful observations and rely on the opportunity to share these for securing maximal reputation benefits.
Both authors and reviewers may comment on such public notes.  

At the beginning of the review process, reviewers are asked to provide a first report. Once all $N_\text{reviewers}$ reports are collected, they are automatically posted (anonymously) in the discussion forum, where they can then be commented on by the public. All reports at this stage follow a standard form for formulating requests to the authors: authors may be asked for
\begin{enumerate}
    \item clarifications, and
    \item specific changes and additions.
\end{enumerate}
The authors' answer is submitted and automatically cross-posted as a reply in the discussion forum. Authors also need to resubmit an improved version of the paper together with their reply. Multiple such rounds may follow: reviewers can either fill out the same form for iterating the discussion, or decide to pass. The process repeats until either \emph{all} reviewers are passing \emph{or} until the authors decide to move forward anyway despite outstanding requests.
In that case, the rounds of iterations end and all actors proceed to the final stage of the review process.
Now, reviewers are asked to provide a more detailed report. This involves to
\begin{enumerate}
    \item score hypotheses as in~\autoref{sec:assessment}
    \item answer how confident they are overall, how much of the paper they have read, and how much of the paper they have understood (these data points may be used to combine the scores)
    \item  write a free-form minimum-length conclusion.
\end{enumerate}  

The scores of the different reviewers are aggregated into a final score by a weighted average (the precise formula is not relevant at this point), and papers above some threshold, to be set by governance, are classified as accepted. The final conclusions are not relevant for the paper acceptance decision; the purpose is for reviewers to provide context for their ratings that might be interesting when evaluating their performance.
Once they are available, the final conclusions and scores for both accepted and rejected papers are visible on a paper result page, but are closed for comments in the forum, which is not intended to create nor influence social consensus about reviewer performance.
Reviewers pick up a remuneration suppressing factor in case their reports are delayed and will eventually be automatically replaced by drawing further reviewers from the volunteers if they exceed a maximum time window to submit their reports.\footnote{If this is not possible, the process stops and the authors are refunded.}

\subsection{Reviewer rating}
This section is concerned with step \ref{th:evaluate} of the process described at the beginning of \autoref{list:steps}.
Everyone with positive reputation can rate reviewers based on the public reviewer reports (on the results page) and the formal reviewer-author communication (in the forum) from one to three stars. This can be done directly on the results page. The average rating of a reviewer is between $1$ and $3$, but is \emph{never} visible. 
The following becomes effective after a threshold of votes are collected (such as 100). For simplicity, we give concrete numbers for the incentives and thresholds, but they could be chosen similarly to \autoref{tab:reputation-gain} to avoid gaming the system.   
Whoever voted ``one star'', when a reviewer's average rating is above $2.5$, loses 5 reputation points. No further benefits can be obtained from a one-star vote, since cancelling the reputation of other reviewers is already somewhat advantageous for the voter;
voting ``two stars'' yields 5 reputation points if the average grade is between $1.5$ and $2.5$ and causes a loss of 5 reputation points if the average is below $1.25$;
voting ``three stars'' yields 5 reputation points if the average is above $2.5$ and causes a loss of 5 reputation points if the average is below $1.5$.
After $100$ votes, with the current average rating score $x$, a reviewer's reputation receives a contribution of 
\[ 100 \log(x-1) \]
which turns positive for $x>2$ and slowly diverges for $x\rightarrow 1$, punishing bad reviews to the point of effectively excluding such reviewers in the future.

%% file: SECTIONS/scientific-impact.tex
\section{Estimating Scientific Impact}
\label{sec:paperquality}
The goal is to create a system that enables open scientific communities to identify and curate good research articles via a binary decision (accept/reject) based on somewhat subjective reviewer polling.  A well-chosen procedure for compressing poll results into this decision increases transparency and decreases subjectivity but must be general enough to accommodate all research works; this procedure is crucial for the success of the system.
This section proposes a score $Q$ for research paper quality based on reviewers rating the importance as well as the strength and conclusiveness of a set of hypotheses that authors have to provide along with their paper submissions. Where possible, rating happens directly on the objectively defined probability scale.

\subsection{Hypotheses}
Similar to~\cite{Curating82:online}, we start by seeing scientific articles as being about hypotheses $H: X\rightarrow Y$ and measure the impact, or quality $Q_H$, of the article regarding $H$ as 
\begin{align} 
Q_H = V_H \cdot L_H,
\end{align}
the product of an overall importance score $V_H$, which captures subject importance and innovation of $H$ assuming $H$ is true, and a learning $L_H$, which quantifies rigor, namely the confidence that is added by the article in $H$ being true.\footnote{We comment on the work that motivates our approach in Appendix~\ref{app:articlequality}.}.
We next develop a notion for $L_H$ using  information theory.

For a hypothesis $H:X\rightarrow Y$, we now consider an outcome $O$ given $X$ that is a Bernoulli random variable either equal to $Y$, or not. This corresponds to saying $H$ is true with probability $Pr(O\!\!=\!\!Y|X)$, or false otherwise.\footnote{Note that, in this subsection, $O$ is not a direct observable but models uncertainty in the truth of a higher-level summary statement $Y$. This notion also applies to theoretical and methodological work. In some cases, statistical features emerging from noisy measurements may be included in $Y$. For example, $Y$ could be a statement about increased \emph{chances} of certain outcomes for a measurement. C.f.~the following subsection~\ref{sec:measurements} for directly referring to measurements instead.}
We propose to use the relative entropy or Kullback-Leibler divergence $D$ and set $L_H \equiv D[ \, P(O|X,A) \, || \, P(O|X) \, ]$ between the distribution of $O$ from before to after considering the article $A$. The relative entropy $D$ measures the information gain as the average reduction of surprise (that is, the reduction of what is unknown about $O$), assuming the distribution of $P(O|X,A)$ is true.\footnote{The surprise (or surprisal), or Shannon information, of an event is the negative logarithm of its probability, such that it is an additive quantity in case of multiple uncorrelated events. The Shannon entropy $S[P_1]$ is the expected surprise, $S  = -\sum p_{1,i} \log p_{1,i}$ in the discrete case. The relative entropy $D[ P_1 || P_2 ]$ is the expectation of the reduction (giving another minus sign) of surprise under $P_1$, so is $D=\sum p_{1,i} (\log p_{1,i} - \log p_{2,i})$.} We suggest to ask reviewers to report their beliefs in the truth of a hypothesis prior to (probability $p$) and after reading the paper (probability $p_A$). Then,
\begin{align}
    L_H &= p_A \log(p_A/p) + (1-p_A) \log((1-p_A)/(1-p)). \label{eq:entropyHypothesis}
\end{align}
If no information is added, $p=p_A$ and $L_H=0$. If $X=Y$, $p=p_A=1$ and $L_H=0$. Furthermore, $L_H>0$ no matter if $p_A>p$ or $p_A<p$. In particular, negating $H$ leads to replacing both these probabilities by their complement, which leaves $L_H$ invariant. We show several more examples in Table~\ref{table:LH}.

\begin{table}[h!]
\centering
\begin{tabular}{||c | c c c c||} 
 \hline
 p & $p_A = 0.1$ & $p_A = 0.7$ & $p_A = 0.9$ & $p_A = 0.98$ \\ [0.5ex] 
 \hline\hline
 0.02 & 0.12 & 3.1 & 4.6 & 5.4 \\ 
 \hline
 0.05 & 0.03 & 2.2 & 3.4 & 4.1 \\
 \hline
 0.5 & 0.53 & 0.12 & 0.53 & 0.86 \\
 \hline
 0.75 & 1.4 & 0.01 & 0.10 & 0.31 \\
 \hline
 0.9 & 2.5 & 0.22 & 0 & 0.07 \\ [1ex] 
 \hline
\end{tabular}
\caption{Added information $L_H$ as given in \autoref{eq:entropyHypothesis} in bits (i.e.~using base-2 logarithm) for various prior beliefs $p$ and updated beliefs $p_A$.}
\label{table:LH}
\end{table}

\subsection{Measurements} \label{sec:measurements}
Some papers report parameter measurements. A hypothesis that is true or false will be inappropriate to describe such contributions. In this case we can see $O$ as a continuous random variable reflecting what is known about that parameter. 
If the measurement result is compatible with what was previously known, we propose to first find the combined result.\footnote{This will usually be sharper. The fact that this is not guaranteed in general (non-Gaussian likelihoods) is a reason for using relative entropy and not the difference of Shannon entropies here, which can be negative if the combined result is broader. In case of Gaussians, the latter would only give the first term of equation~(\ref{eq:entropyMeasurement}). Note that this equation instead also captures a shift of the mean in the second term. If one were to use the new measurement alone and not the combined result for the paper results, very bad measurements $\sigma_A \gg \sigma $ and $|\mu_A - \mu| \sim \sigma_A$ could easily create a large contribution to $L_M$ from this term. Thus, for this term to make sense, it is indeed the combined result that must be used here.} 
For Gaussians $\mathcal{N}$ with means $\mu$ and standard deviations $\sigma$, the measurement learning $L_M$ is
\begin{align}
    L_M \equiv D[\mathcal{N}_A \, || \, \mathcal{N}] = \log (\sigma/\sigma_A) + \frac{\sigma_A^2 + (\mu_A - \mu)^2}{2 \sigma^2} - \frac{1}{2} \label{eq:entropyMeasurement}
\end{align}
where the subscript $A$ refers to the Gaussian after including the new measurements.
It is possible that the reviewer has some doubts that the new result holds and assigns a probability $P_A<1$ to this case. One could then consider the mixture model between the new, combined result with weight $P_A$ and the old result with weight $1-P_A$, but this does not give a closed-form expression for $L_M$. A simple heuristic would be to interpolate $\mu_A$ and $\sigma_A$ towards their old values with decreasing $P_A$. 

If, instead, the new measurement is incompatible, or if it is fully compatible but the novelty lies in reducing a previous risk, the suggestion to the authors is to instead phrase their result as a hypothesis.

\subsection{Total assessment} \label{sec:assessment}
Often, papers will make a series of claims of varying nature. For independent statements, one would want learning scores to be additive. The (relative) entropy on product distributions has this property. However, we do allow for weighting the statements. This  decomposes an overall importance $V$ into $V_i = w_i V$, where $\sum_i w_i = 1$ and $i$ iterates through all hypotheses and measurements. We end up with the following formula for evaluating the research paper,
\begin{align}
\label{eq:paperquality}
    Q = V \sum_i w_i L_i  
\end{align}
where for $L_i$ we use~(\ref{eq:entropyHypothesis}) for statements and~(\ref{eq:entropyMeasurement}) in case of new, improved measurements. The overall importance $V$ is chosen on some discrete scale by each reviewer and quantifies the progress in the field \emph{under the assumption that all hypotheses are true}.\footnote{$V$ could be chosen on the scale \emph{unimportant}, \emph{incremental improvement}, \emph{generalization}, \emph{substantial improvement}, \emph{discovery}, \emph{breakthrough}. The actual delivery of the paper with regards to its claims is then quantified by $L_H$ or $L_M$.}

For the purpose of this work, we attempt to design a peer review system that allows to measure the paper quality~(\ref{eq:paperquality}) as accurately and objectively as possible.
In practice, on submission of their work, authors must list the statements or measurements together with their relative importance weights $w_i$ (and input the measurement errors $\sigma$ and $\sigma_A$ and means if applicable).
Here, weights need to be larger than a minimum, $w_i > w_\text{min}$, which limits the number of statements that can be submitted.
The reviewers then choose the overall importance $V$,
and their probabilities of prior and posterior truth $p_{i}$ and $p_{i,A}$, or $P_{i,A}$, respectively, for each of these statements or measurements. 

The list of weighted statements is expected to be a particularly powerful summary of a research paper, more so than a colorful abstract.
This is because authors are incentivized to minimize the list to those statements that are well-supported by the paper.
Otherwise, assuming that more statements raise the value of $V$ that a reviewer would assign only insignificantly,  the quality score decreases when confidence is ranked low for the weight fraction of overall $V$ corresponding to less convincing statements.
The $Q$ values for different reviewers are aggregated into a final score, and papers above a threshold are classified as accepted. 

%% file: SECTIONS/tokenomics.tex
\section{Tokens and Incentives}
\label{sec:tokenomics}
The review process is backed by a tokenized reward system providing both monetary and non-monetary incentives for users.
We employ three (fungible) tokens: a reputation token, a ``science'' token and a stablecoin. These tokens interact with different groups: authors (submit papers), reviewers (curate papers), the protocol community (govern) and the wider community (investment/trading). 

The reputation token is nontransferable and tracks user accomplishments. It helps establish a user's reputation within the platform based on their contributions and activities.
The "science" token, on the other hand, is a transferable token that represents project shares. It can be traded on exchanges and has fee sharing and governance utility on the platform.
Finally, a stablecoin is provided externally and serves as a more convenient means of payment within the platform.

The peer review of scientific work stands in close analogy to the curation process on the art market. Simultaneously, papers and various other contributions on the platform are ideally represented by tamper-proof human-readable certificates beyond a single reputation number. This suggests employing also non-fungible tokens (NFTs) as 
\begin{enumerate}
    \item a paper NFT, created upon acceptance after review as part of the collection of accepted papers, to be sold (by the protocol) and traded on external NFT markets, and further:
    \item preprint NFTs, proving submission of a preprint at some time, 
    \item author NFTs, each proving one of the holder's papers has been accepted, 
    \item review NFTs, each proving that a user has carried out a specific review,
    \item voucher NFTs, each enabling holders to request a review. 
\end{enumerate}
Only the paper NFT is transferable, all other certificate-like NFTs are not.

Authors are required to pay a fee in stablecoin or use a voucher when submitting a paper for review~(see \autoref{sec:estimations} and \autoref{subsec:business-model}).
Upon acceptance of their work, a paper NFT is minted and auctioned off by the protocol, and authors receive reputation.
Reviewers are remunerated either in a stablecoin payout or voucher for their work, e.g. an equivalent of \$750~(see \autoref{subsec:business-model}).
They also receive (or lose) reputation when they participate in the different stages of the review process, as discussed in~\autoref{sec:review}.
Investors can buy the science token and stake it to receive an effective interest rate on their investment.
These ``real'' yields are part of the platform revenue.

Additionally, reputation allows ``farming'' of inflationary science token rewards by staking it, so that authors and reviewers have additional incentives to reach higher reputation, and are invested in the success of the project. This is an implementation of the points/coin two-token model suggested in~\cite{repsystems}.
Besides the two forms of staking the science token, the token is also used for governance by active members (authors and reviewers) to modify system parameters (see \autoref{subsec:governance}).

\subsection{Reputation}
\subsubsection{Reputation as a dynamic quantity}
The state of the art in a given area of research often evolves significantly over the span of a career. Designing reputation as a dynamic quantity including decay takes this into account; new talents get a fair chance to earn their place in the community when established ones get less involved. Similar ideas have been voiced in~\cite{repsystems}.
Let the time evolution of a user's reputation $r(t)$ be governed by the differential equation
\begin{align}\label{eq:reputation-decay}
    \frac{d}{d t} r &= -k(t)\times r + f(t),
\end{align}
where $k(t)$ sets the reputation decay and is updated by governance and $f(t)$ is a source term to be interpreted as the productivty of a specifc user.
While in this representation as a differential equation a sudden jump of reputation at some time technically corresponds to a weighted Dirac delta contribution to $f(t)$, for long time-scales (of the order of the inverse of the rate $k$) $f(t)$ may be assumed to be smoother and can be modelled as a time-averaged reputation gain ($f>0$) or loss ($f<0$) due to that user's actions.

\begin{figure}
    \centering
    \includegraphics[width=.7\textwidth]{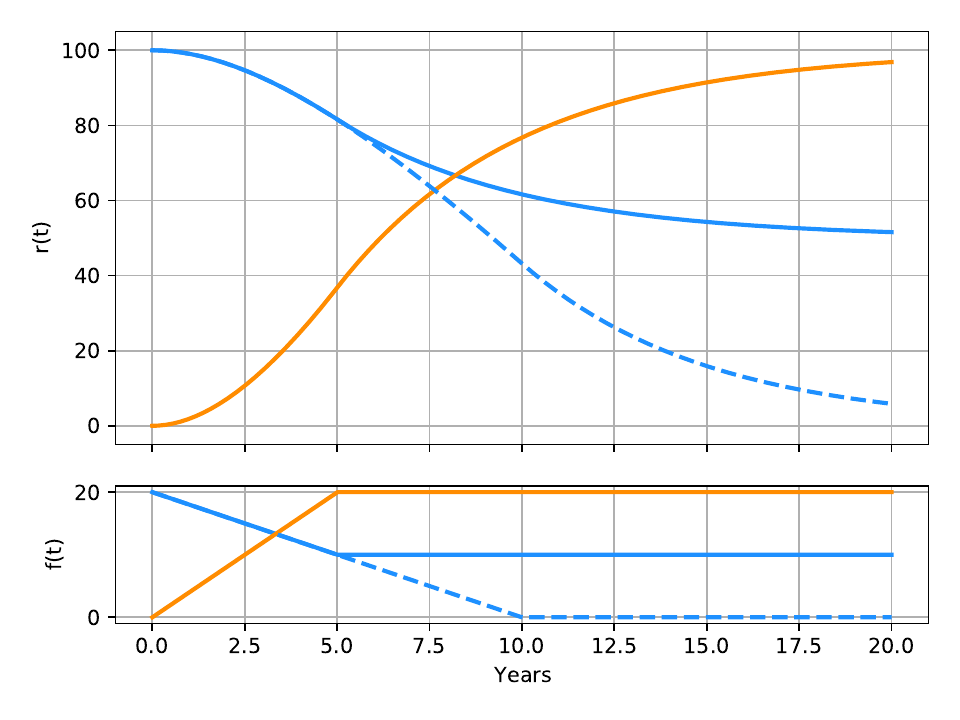}
    \caption{Reputation decay according to \autoref{eq:reputation-decay} for two scientists, one senior at the end of their career with high initial equilibrium reputation $r=100$ but with decreasing activity $f(t)$ (blue) either to half of the initial $f(0)=20$ over 5 years (solid), or to zero over 10 years (dashed), and one, junior, starting at zero reputation and increasing their activity to the same level $f=20$  level as the senior had initially over a period of 5 years. Here, $k=1/5y$. Thanks to reputation decay it only takes a few years after their contribution rate crosses for reputation to reflect this instead of a full career.
    }
    \label{fig:repdecay}
\end{figure}

When $f(t)$ of a user takes on a new value, such as when they start using the system more or retire, and $k(t)=k$ is constant, $r$ decays exponentially to the ratio of these two rates on a time-scale given by $1/k$. As shown in Figure~(\ref{fig:repdecay}), this allows smooth yet faster adjustments compared to a system without any decay, where the newcomer would only at the end of their own career catch up with an equivalently strong retiree.

The decay rates should operate on quite large time scales resembling the loss of up-to-date expert knowledge.
The half-life period could initially be set to 5 years.
Together with the reward and slashing mechanisms of the submission, review or recommendation modules, reputation decay encourages active engagement.

\subsubsection{Recommendations}
It is a common problem that long-term members aggregate more voting power simply due to their early-stage commitment.
New members have a natural disadvantage when entering the system before the system fully captures their expertise (on times shorter than the decay time), which may lead to an unjust dominance of early adopters.

We therefore propose to allow members with sufficient reputation to \emph{recommend} new or existing members for certain tasks.
A correct forecast, i.e. that the recommended user becomes a valuable part of the community, can be rewarded by a reputation gain for both.
The reward peaks for the correct estimate and declines both for over- and under-estimation; overestimation is penalized harder to incentivize conservative estimates rather than high-risk predictions.

\subsection{Governance and staking}\label{subsec:governance}
Decentralized projects that are not immutable protocols need decentralized governance. Typically, voting power is directly proportional to the amount of some goverance token that a user has to buy and stake. This is a plutocratic form of governance, but it is sybil-resistant, that is, there is no advantage from creating multiple accounts and splitting the stakes. 

Another interesting idea, which however requires protection from sybil attacks, is quadratic voting~\cite{lalley2018quadratic}, whereby the price of each additional unit of voting power grows linearly with the amount of voting power someone has already acquired, such that the total cost of the vote is quadratic in the total power. This price increase can stop wealthy individuals who strongly benefit economically from a certain outcome from swinging the vote all the way in their favor, even when we assume the chance of this outcome being realized increases linearly with their voting power.\footnote{They would buy only as much voting power until the price of an additional unit exceeds its expected economic benefit, which is assumed to be constant here.}

Reputation $r$ is, arguably, a good alternative proxy for voting power $P_\text{vote}$ instead of monetary assets, and causes some sybil resilance; supplementing a proof of identity is still an advisable goal.\footnote{The amount of reputation itself forms, due to its non-transferable nature, a part of the identity of a user. However, it would be possible to keep multiple accounts from the beginning, and excess reputation can be gained with nearly no extra work from the voting-based reputation incentives by voting the same on all accounts, and, with additional work, due to increased chance of being a reviewer on any of these accounts. Only \emph{large} amounts of reputation stemming from many reviews would alone signify the existence of another real person (or multiple) behind the account.} We consider a scheme similar to quadratic voting, that is,
\begin{equation}
    P_\text{vote}\propto r^\frac{1}{2}.
\end{equation}
While the economic arguments of~\cite{lalley2018quadratic}, as summarized above, do not apply, a power law with exponent $<1$ helps making the governance process more egalitarian in a context where many new users enter the system at any time. 

However, voting power should not only depend on the expertise, but also on the \emph{commitment} of a user.
Commitment to the platform requires a user to stake an amount of science tokens in relation to her reputation.
By committing to the platform, a user does not only risk the value of her reputation in case of governance decisions that are unfavorable for the protocol, but also monetary value.
To this end we introduce a \emph{commitment factor}  $c$ into the voting power,
\begin{equation}
    P_\text{vote} = c \times \sqrt{r},
\end{equation}
which is given by the bounded ratio of staked science tokens $s$ and the reputation held by a user:
\begin{align}
    c &= \min \left(1,\frac{s}{\alpha \,r}\right),
\end{align}
such that the voting power can maximally reach $\sqrt{r}$ by increasing $s$. Here, $\alpha$ is a constant with units of science token per reputation that can itself be adjusted by governance. Hence, governance power can only be obtained via reputation together with a certain amount of staked science token. For sufficiently small $\alpha$, the required amount of money per reputation gained from review can be smaller than the corresponding reviewer payouts (this requires regularly adjusting $\alpha$ to changes in the dollar price of science tokens). Furthermore, users that currently max out their governance power ($c=1$) who gain reputation immediately obtain a bonus of additional science tokens $s\propto \sqrt{r}$ such that their $P_\text{vote}$ is at least held constant (after which yet $c<1$, that is, $P_\text{vote}$ is not maximal, such that topping up $s$ with own funds for $c=1$ is still required for getting another bonus for the next reputation increase).\footnote{Note that as a function of changing reputation, the science tokens required to maximize $P_\text{vote}$ behave like in quadratic voting as well in this system: $s_\text{max} = \alpha r = \alpha P_\text{vote}^2$.}
Considerable staking periods, e.g. $T_\text{stake}=4\text{y}$ incentivizes a long-term commitment to the platform and discourages short-sighted and harmful decisions by individuals.

Besides the mentioned bonus, users earn an high inflationary interest rate on part of their staked science tokens. 
The commitment factor cannot be larger than one to prevent people from simply buying voting power, 
and, similarly, users only earn an interest rate on at most $s_\text{max} =  \alpha r$ science tokens (for which $c = 1$). Engaged users are able to increase their earnings by actively participating in the system and increasing $s_\text{max}$ with their reputation. If a large fraction of the token demand is coming from \emph{investors} (instead of users with reputation), overall token dilution can be small even when these reputable \emph{user} rewards are large. Further rewards for staked science tokens \emph{exceeding} $s_\text{max}$ can be made available but are restricted to ``real yield'' and are available to all investors, that is, stablecoin protocol profits that can be partially distributed to stakers.

Users can change the system's parameters via \emph{petitions}.
Petitions are public proposals, which users can support by voting leveraging their voting power $P_\text{vote}$.

\subsection{Business model}\label{subsec:business-model}
Monetary remuneration of reviewers requires a source of income for the protocol.
We identify four such potential revenue streams.
First, authors paying for review provides the most direct means of financing a review.

Second, a premium subscription could provide useful personalized content and conveniences, e.g.~a personalized digest of interesting articles, suggested contacts or collaborators, relevant conferences and more.
Additional services for collaboration may also be eventually offered. A premium membership would also signal support of the platform and could come with voluntary donations that can be included in a public ranking.

The third income stream is web3-native. As mentioned, paper NFTs attesting successful publication of a scientific work will be auctioned off by the protocol. Institutional entities, DAOs or private investors might seek to support and invest into science in general and certain scientists in particular. However non-experts need to rely on a strict curation process that vouches for what they are buying, with two effects: First, the work represented by an accepted paper NFT is likely to be good science. Second, the protocol ensures provenance once it is established as a rigorous journal. The concept of provenance and authenticity is one that correlates strongly with value in the art markets. This is similar to art investors trusting art experts in their judgement of both quality and authenticity. Strict peer review is a scientific curation process, and thus enables this novel avenue of funding science. The downside is that the returns cannot be estimated and will vary depending on how much a paper captures the public interest and other market effects.

Fourth and finally, the DAO could decide to sell part of the science token treasury to fund operations.

Authors should not be rewarded with a fixed fraction of the paper NFT auction amount and/or future NFT trading commissions. The public opinion is not a suitable judgement call for the scientific importance of work in various areas, and authors should not be incentivized to optimize their interests for higher profits. A moderate cap on the absolute amounts authors can earn solves most of these issues while allowing for additional reasons for authors to choose to submit their papers to the platform and for investors to still support particular scientists.

As a more direct means to structure payments around the peer review process, we suggest a cooperative model, where reviewers can choose to receive a voucher instead of a direct payout for their work, which they can use to get a review for their own work, or donate to whitelisted (pre-selected) addresses.\footnote{A free, competitive and global market for vouchers could lead to devaluation of reviews, and selling/transfers of vouchers should potentially be disallowed.}
Pay-outs (or cost of review requests) without a voucher would be reduced (or increased) in value to encourage use of vouchers and therefore engagement as both, reviewer and author.
This model would lower the means for extensive funding and reduce taxes.
For more details, we refer to \autoref{sec:estimations}.

We stress that the platform serves first and foremost the scientific community.
All fees and subscriptions should be designed in a transparent and ethically responsible way - and blockchain technology is very useful to this end.

\subsection{Estimations}\label{sec:estimations}
To estimate number of publications and required funding we orient this case at Nature publishing.
Nature receives more than 10k submissions/year with an immediate rejection rate of about 60\%. The rate of acceptance is about 5\% of submissions~\cite{NatureSubmissionStats}.
Assuming about 10k submissions/year, an immediate rejection rate of 70\% before review and three reviews per paper, this results in 9k required reviews with a  cost of \$6.750.000 when targeting an average cost of \$750 per review.
Aiming for a final acceptance rate of 5\% with respect to initial submissions, this results in 500 accepted papers/year.

Denoting the base value of a review by $x$, a natural starting point is to require a payment $x$ per review from the authors.
In order to reduce the submission costs for the authors, we introduce a discount $d$. Ideally, $d=x$, s.t. submissions are free of charge.
The remuneration of the reviewers should be unaffected by the author discount $d$.
We are able to set $d>0$ if the protocol is funded by the income streams not related to the review system (c.f.~second, third and fourth point in the previous section).
An author can either use a voucher to request a review or pay up-front. 
Along the same lines, a reviewer can decide to receive a voucher or a cash payout.
In order to incentivize the use of vouchers and therefore the mutual exchange of reviews, we can introduce a penalty $\Delta>0$ for using cash.
Then, review requests via cash cost $x-d+\Delta$, while cash payouts for reviewers are reduced to $x-\Delta$.
Reviewers accepting a voucher receive an additional cash payout of $d$. This mechanism would benefit active members of the scientific community that are both authors and reviewers; in particular authors would be incentivized to review to fund their submissions and reviewers are encouraged to submit their papers with the protocol (without concerns about tax reporting arising if $d=0$ or a reviewer chooses to donate the extra payout $d$ besides the voucher).
Furthermore, it is possible to introduce unbacked vouchers into the system,~e.g. by initially handing them out to early supporters, as long as in case of insufficient cash funding of the protocol, reviewers are informed that they can only review against a voucher reward instead of cash payout. When the protocol earns from e.g.~NFT sales it can eventually transition into a fully backed state and/or allow for a discount $d>0$.

Let us also consider an intermediate cost tier for papers that are reviewed but rejected.
Since those cases still cause the same work for the reviewers, but do not result in an accepted paper, a reduced fee seems reasonable. 
This can be achieved by introducing another discount $d_R \ge d$.\footnote{In practice, this would be realized by working with $d$ and refunding $d_R - d$ (times e.g.~three reviews) in case of rejection.}

Finally, let us consider the  particular case of remunerating the reviewers solely from accepted papers, that is, when rejected papers are free, $d_R = x$. 
Since 
\begin{equation*}
    3\times 750\$ \times \frac{30\%}{5\%} = 13.500\$,
\end{equation*}
the cost to finance all reviews by the ones that are actually published is at most of the same order of magnitude as the submission with Nature\footnote{Nature charges up to 11.390\$ to publish a single open-access article~\cite{NatureFees}}, with the notable difference being that reviewers receive a proper payment for their work.
Furthermore, if accepted paper NFT sales yield an average value of 13.500\$, which is not unrealistic, we can achieve $d=x$, that is, all reviews are free. As mentioned, further protocol profits can also be distributed to all science token stakers at the discretion of protocol governance, which helps to create wider demand for the science token and backs up the value of the inflationary token rewards users with reputation can receive. Profits from an initial token sale could be used to pre-fund and kickstart the protocol with free reviews.

The system could offer more options for reviewers to (anonymously) donate the value they generate or to accumulate value into a grant in order to support larger fixed-cost expenses such as the salary of a graduate student for a certain amount of time.

%% file: SECTIONS/conclusion.tex
\section{Conclusion}
\label{sec:conclusion}
We identified missing incentives as a key problem for peer review, lowering the quality of reviews and slowing down the publication process.
By quantitatively assessing the scientific impact of a paper applying information theory to plausibility estimates of author hypotheses, and creating a review mechanism, we aim to reward high-effort reviews with monetary and non-monetary incentives while keeping publication costs low.
The proposed decentralized system for open peer review puts the responsibility in academic publishing back into the hands of the scientific community.
Based on the mutual exchange of time and effort between authors and reviewers and further sustained via a \emph{pay-per-review} business model, potential NFT-sales (in analogy with curation on the art market) as well as investments, the proposed system promises to increase publication speed and quality while preventing scientific misconduct (like biases, plagiarism, control or censorship) via transparent peer review and reputation tracking.

%% file: SECTIONS/ack.tex
\section{Acknowledgements}
The authors thank Henrik von der Emde, Jan Ole Ernst, Fei Ding, Phillip Koellinger, Dusan Kolarski, Steffen Sahl, Marko Simonovic, Stefan Ulmer and Zvonimir Vlah for fruitful discussions.

%% file: SECTIONS/appendix.tex
\section{Appendix}\label{sec:appendix}
\subsection{Comment on previous work on article quality} \label{app:articlequality}
Reference~\cite{Curating82:online} introduces a measure of added information $I$ for studies estimating the impact of $X$ on achieving that the outcome $O$ of the experiment is indeed $O=Y$, $$1+I = Pr(Y|X)/Pr(Y) = Pr(X|Y)/Pr(X).$$ However, for $L_H$ they choose the bounded expression $L_\Delta = Pr(Y|X) - Pr(Y)$. $L_\Delta$ can be large just because $X$ is similar to $Y$: Uninteresting experiments that test $H_\text{trivial}: X \rightarrow Y \sim X$ by choosing to set $Y$ close to the condition $X$ (or vice versa) default to higher learning $L_\Delta$, which would need to be compensated by a decline in $V_H$.
$L_H$ itself should be zero for $X=Y$, since nothing new can be learned about a trivial statement.
$L_\Delta$ also depends too strongly on an overall rescaling factor $f$ of both $Pr(Y)$ and $Pr(Y|X)$ instead of measuring the information the paper \emph{adds}.
The expression $I$ does not suffer from this issue for fixed $X$, but it is still unbounded in the limit in which first $X$ becomes $Y$ ($H \rightarrow H_\text{trivial}$) and second the common factor $f$ is taken to zero, which is unacceptable.

\subsection{Other efforts}
The following serves to provide a brief overview of other desci projects and efforts directed towards improved peer review.

\paragraph{AntsReview}
Smart contracts termed \emph{AntsReview} enable researchers to install agreements over the review of a scientific work and potential remuneration of the participating reviewer~\cite{AntsReview}.
The AntsReview contract allows the author of a paper to determine a bounty for the review, which will be payed to the reviewer after the review has been scrutinized and the contracted marked as fulfilled by a third party community member.

\paragraph{Atoms}
\emph{Smart research contracts} shall serve to create a community driven instrument to direct funds to projects and researchers~\cite{Atoms:online}.
This way, researchers can receive financial and reputational incentives for their research.
The smart research contracts are mediated by a decentralized peer-to-peer review network that decides which projects or researchers are to be funded.

\paragraph{ResearchHub}
As a \enquote{GitHub for Science}, \emph{ResearchHub} aims to provide a platform for open science by sharing open-access publications, discussing research and rewarding contributions to the scientific community with a token called \emph{ResearchCoin}~\cite{ResearchHub:online}.

\paragraph{DeSci Foundation}
The foundation helps to advance high-quality research by awarding grants to researchers and encourages the development of web3-tools for science~\cite{DeSciFoundation}.
The \emph{DeSci Labs} develop a full web3-based ecosystem for decentralized science~\cite{DeSciLabs}.
Integral elements of this ecosystem are \emph{DeSci Nodes} that serve to store, access and interact with research artefacts and Autonomous Research Communities, so-called \emph{Arcs}.

\paragraph{ScienceFund}
As a DAO, \emph{ScienceFund} develops web3 protocols in order to coordinate community efforts to fund science~\cite{ScienceFund}.
It promises to transform science donations into strategic investments, the impact thereof traceable via the Science Funding Token.
Acceleration funding is proposed as a application-free grant that is targeted at high-performing research projects in order to speed up their progress.

\paragraph{DeSci World}
The DeSci-Dashboard of DeSci World provides an overview of the multiple desci efforts that happen around the world.
They provide a platform for the community to connect and set their goal to decentralize all scientific institutions~\cite{DeSciWorld}.

\paragraph{OpSci}
As an autonomous open research community, OpSci employs web3 tools to make science more accessible, enhance data availability, democratize funding and enable better collaborations~\cite{Opsci}.

\paragraph{VitaDAO}
A decentralized funding entity for research in the biotech sector.
VitaDAO has already community-funded two research projects.
Its mission is to \enquote{extend the human lifespan by researching, financing, and commercializing longevity therapeutics in an open and democratic manner.}~\cite{VitaDAO}

\paragraph{SciNet}
As an investment platform in the life sciences, \emph{SciNet} aims at connecting institutional investors with researchers in order to directly fund research projects~\cite{SciNet}.
This process is steered and secured by web3 protocols, which ensures the authenticity of results and protection of intellectual property rights.

\paragraph{Review Commons}
\emph{Review Commons is a platform for high-quality journal-independent peer review in the life sciences}.
Authors can submit their work to Review Commons, where it undergoes a peer-review process and is then published as a pre-print as well as forwarded to a journal of choice~\cite{ReviewCommons}.